\documentclass[11pt]{article}

\title{Average-Case Parsing of Unambiguous Context-Free Grammars}
\author{Kazuhiro Inaba \\ \texttt{kiki@kmonos.net}}

\usepackage{amsmath,amssymb,amsthm}
\usepackage{xcolor}
\usepackage{url}

\newtheorem{theorem}{Theorem}
\newtheorem{lemma}[theorem]{Lemma}
\newcommand{\N}{\mathcal{N}}
\newcommand{\wt}{\operatorname{wt}}
\newcommand{\Search}{\mathsf{Search}}
\newcommand{\fail}{\mathsf{fail}}
\newcommand{\eps}{\varepsilon}

\begin{document}

%%%%%%%%%%%%%%%%%%%%%%%%%%%%%%%%%%%%%%%%%%%%%%%%%%%%%%%%%%%%%%%%%%%%%%%%%%%
\maketitle

\begin{abstract}
We show that every fixed unambiguous context-free grammar admits a
deterministic parser with expected running time $O(n^{1.5})$ on a
uniformly random word of length $n$. The parser utilizes a certain
form of sparsity estimate supplied by the unambiguity itself.
\end{abstract}

%%%%%%%%%%%%%%%%%%%%%%%%%%%%%%%%%%%%%%%%%%%%%%%%%%%%%%%%%%%%%%%%%%%%%%%%%%%

\section{Introduction} \label{sec:intro}

The asymptotically fastest known algorithm for recognizing a general
context-free grammar runs in $O(n^\omega)$ time via Valiant's
reduction~\cite{Valiant75a} to matrix multiplication.
When the grammar is unambiguous, i.e., each word has at most one parse
tree, the bound can be improved. Earley's parser~\cite{Earley70} runs in
$O(n^2)$ worst-case time for a fixed unambiguous grammar. Manacher's
variant of the Cocke--Younger--Kasami algorithm refines this bound in
terms of the number of derivable subwords~\cite{Manacher78}, although
the worst case stays quadratic.

To our knowledge, however, there is no specific unambiguous grammar for
which a linear-time parsing algorithm is unknown.
Well-known examples of unambiguous languages are 
even-length palindromes without center markers
$\{ww^R \mid w \in \{\mathtt{a},\mathtt{b}\}^*\}$ and the disjoint union
of counting languages $\{\mathtt{a}^n\mathtt{b}^n \mid n \geq 1\} \cup
\{\mathtt{a}^n\mathtt{b}^{2n} \mid n \geq 1\}$,
both being trivially linear-time parsable.

There are also more elaborate unambiguous languages.
Write the input word as $w_1\#\cdots\#w_k$, with $k$ odd and
$w_i\in\{\mathtt{a},\mathtt{b}\}^*$. Assign weights $\wt(\mathtt{a})=1$
and $\wt(\mathtt{b})=2$, and extend the weight by addition. Accept if
there are cuts $w_i=u_iv_i$ satisfying
$\wt(v_i)=|u_{i+1}|$ for all $1\le i\le k$ with $u_{k+1}=u_1$.
For example, $\mathtt{aaab\#aba\#bab}$ has the cuts
$\mathtt{aaa|b\;\#\;ab|a\;\#\;b|ab}$.
For any
proposed prefix length $t_i$, the next length is the suffix weight
$\wt(w_i[t_i:])$, a strictly decreasing function on its domain.
Two distinct feasible initial lengths would reverse their order after
every step, hence after an odd number of steps could not both return
to themselves. The unique cuts determine every production.
Hence, the grammar is unambiguous.
Although it is less obvious than in the previous examples, this language
still admits linear-time parsing. Tabulate all suffix
weights, try each cut of a shortest block, and propagate around the cycle.
Discard a candidate as soon as its next cut is outside the next block. If
the shortest block has length $m$, the work is $O(n+k(m+1))=O(n)$,
since $k(m+1)\le n+1$.

Zimmerman~\cite{Zimmerman22} raised the possibility that every fixed
unambiguous grammar admits a worst-case linear-time parser. Although
this question still remains open, our paper takes one step forward in this
research direction. By looking into the \emph{average} time complexity,
we show that any unambiguous context-free language can be parsed
in subquadratic, expected $O(n^{1.5})$ time, on a uniformly random word
of length $n$ including words outside the language.
The key observation underlying our algorithm is that, if an unambiguous
grammar has a rule $A\to BC$, either $B$ or $C$ must denote a relatively
sparse language. Equivalently, if both were dense,
the concatenated language would inevitably admit multiple parse trees.
The sparser side can be parsed by Manacher's algorithm without modification.
Guided by the chart computed for the sparser side, the recursive search
on the denser side is shown to terminate in linear time on average.

%%%%%%%%%%%%%%%%%%%%%%%%%%%%%%%%%%%%%%%%%%%%%%%%%%%%%%%%%%%%%%%%%%%%%%%%%%%

\paragraph{Related Work.}
Use of sparsity to accelerate the parsing time is not new.
Manacher's algorithm~\cite{Manacher78}, which we critically rely on, was
already based on this idea. Bernardy and Claessen~\cite{BC15} showed
that under a certain sparsity assumption, a modified version of Valiant's
algorithm for general context-free grammars can run in $O(n)$ total work.
In our algorithm, sparsity is not an explicit assumption. Rather, we
derive the sparsity necessary for parsing from the unambiguity assumption.

Cohen and Roth~\cite{CR78} analyzed the average-case time complexity for
deterministic parsers. They average over words of a given
length generated by the grammar, while our sample space is
all the words of the same length, including words outside the language.

%%%%%%%%%%%%%%%%%%%%%%%%%%%%%%%%%%%%%%%%%%%%%%%%%%%%%%%%%%%%%%%%%%%%%%%%%%%

\section{Grammars, Uniform Inputs, and Charts} \label{sec:prelim}

A \emph{context-free grammar} is a quadruple
$G=(\N,\Sigma,R,S)$, where $\N$ is a finite set of nonterminals,
$\Sigma$ a finite set of terminals, $R$ a finite set of rules,
and $S \in \N$ a start symbol. Throughout the paper, we assume
Chomsky normal form (CNF), where each rule is $A\to BC$ or $A\to \mathtt{a}$,
where $A,B,C\in\N$ and $\mathtt{a}\in\Sigma$.
An optional $S\to\eps$ is allowed when $S$ occurs on no right-hand side.
For $\alpha = \beta A \gamma$, if there is a rule $A \to \delta$,
we write $\alpha \Rightarrow \beta\delta\gamma$. The reflexive transitive
closure of $\Rightarrow$ is denoted by $\Rightarrow^*$.
For a nonterminal $X$, we write its language as
$L_X = \{w \in \Sigma^* \mid X \Rightarrow^* w\}$.
A parse tree for an input word $w \in \Sigma^*$ is an ordered tree labeled
by $R$, such that the root is labeled by a rule with $S$ on the left-hand side,
the two children of a node labeled $A \to BC$ are labeled by $B$- and $C$-rules,
the leaves are labeled by rules of the form $A \to \mathtt{a}$,
and the yield of the tree is the terminal word $w$.
Parse trees for $w$ are in one-to-one correspondence with the leftmost
derivations of $S \Rightarrow^* w$.

The grammar is \emph{unambiguous} if each word has at most one parse
tree from $S$. An unambiguous language is one admitting such a grammar.
CNF conversion can preserve unambiguity. Handle empty input separately, delete
the empty production, and trim unreachable and unproductive symbols. An
unproductive start symbol rejects all nonempty words.
Otherwise every remaining nonterminal is itself unambiguous,
since two trees from $X$ could be inserted into a fixed terminal context for $X$
at the start symbol.

Let $s=|\Sigma|\ge 1$ and let $W_n$ be the uniform distribution on $\Sigma^n$.
Its letters are independent and uniform, so the probability of $u$ matching at
a fixed valid position is $p(u)=1/s^{|u|}$. Throughout this paper, all
expectations are with respect to this input distribution.
We use a word RAM with $\Theta(\log(n+2))$-bit words. The grammar is fixed and can be
preprocessed once. All asymptotic constants may depend on $G$.

For an input $w$ of length $n$, a \emph{chart entry} is a triple $(X,i,j)$
such that $0\le i<j\le n$ and $w[i:j]\in L_X$. Here $[i,j)$ is a half-open
span. The \emph{chart rooted at} $Y$ contains all such entries whose label $X$ is
reachable from $Y$, including $Y$ itself. Manacher's seminal work demonstrated
that the parsing time for unambiguous grammars can be bounded by the size of the chart.

\begin{lemma}[Manacher's chart algorithm {\cite{Manacher78}}]\label{lem:manacher}
For any nonterminal $Y$ and input $w$, let $F$ be the number of entries in
the chart rooted at $Y$. The entire chart can be listed in $O(n+F)$ time
and space, including on inputs outside $L_Y$. Each entry can carry its
rule and, for a binary rule, pointers to its two child entries.
\end{lemma}

Note that Manacher's algorithm alone does not give the desired average-case
bound. There are unambiguous grammars for which any input word has
$\Theta(n^2)$ chart entries.

%%%%%%%%%%%%%%%%%%%%%%%%%%%%%%%%%%%%%%%%%%%%%%%%%%%%%%%%%%%%%%%%%%%%%%%%%%%

\section{Selective Charting with a Mass Oracle}\label{sec:selective}

Suppose we are parsing a span with a rule $A\to BC$. If we have a chart
for $B$, each matching prefix supplies a candidate split, and we need only
parse the remaining suffix recursively from $C$. A chart for $C$ lets us
do the same from the other end.
By Lemma~\ref{lem:manacher}, fewer successful matching subwords imply faster
chart construction. In this section, we show an algorithm assuming a unit-cost
oracle that determines which child produces a smaller chart.
How to remove the oracle and complete the full algorithm will be explained in
the next section.

The \emph{mass} of $X \in \N$ up to length $n$ is
\[
 M_X(n)=\sum_{k=0}^{n}\frac{|L_X\cap\Sigma^k|}{s^k}.
\]
Each summand is the probability that a uniform length-$k$ word belongs to
$L_X$. Suppose an oracle compares masses at unit cost. For each binary
rule, choose the lower-mass child as its \emph{charted child},
breaking ties to the left. Recurse on its other child, the
\emph{search child}. Use this choice at every occurrence of the rule.

\paragraph{Algorithm.}
In the algorithm description below, ``Ensure the $X$-chart exists'' means
running Manacher's algorithm for the nonterminal $X$ for the whole input word
if the chart has not yet been computed. The chart is indexed by the start and
the end positions, ordered by length. This takes $O(n+F)$ time.
Calling $\Search(S,0,n)$ on the full half-open span either returns the parse
tree for the whole input word or rejects it.
\begin{quote}\fbox{\begin{minipage}{\linewidth}\small
\noindent\textbf{Search}$(A,i,j)$\\
\hspace*{1em}If $j-i=1$, return the leaf for $A\to w[i]$ if present, else $\fail$.\\
\hspace*{1em}For each binary rule $r=(A\to BC)$:\\
\hspace*{2em}  If $B$ is the designated "smaller" child:\\
\hspace*{3em}    Ensure the $B$-chart exists.\\
\hspace*{3em}    For each $B$-entry $[i,k)$ starting at $i$ with $k < j$:\\
\hspace*{4em}      Let $t = \Search(C,k,j)$. If it succeeds:\\
\hspace*{5em}        return $r$-labeled tree with $(B,i,k)$'s tree and $t$ as children.\\
\hspace*{2em}  Otherwise:\\
\hspace*{3em}    Ensure the $C$-chart exists.\\
\hspace*{3em}    For each $C$-entry $[k,j)$ ending at $j$ with $i < k$:\\
\hspace*{4em}      Let $t = \Search(B,i,k)$. If it succeeds:\\
\hspace*{5em}        return $r$-labeled tree with $t$ and $(C,k,j)$'s tree as children.\\
\hspace*{1em}Return $\fail$.
\end{minipage}}\end{quote}
The algorithm tries every valid split, and each recursive call tests
a strictly shorter span. Induction on span length gives correctness for
\emph{any} fixed choice of side per rule.

The following lemma ensures that chart construction is inexpensive for
at least one of the right-hand side nonterminals in each rule $A\to BC$.
In particular, we can prove the mass of the smaller child is of order $O(\sqrt{n})$.
It implies that within the set of subranges of a length-$n$ input word,
on average, the designated charts contain $O(n \sqrt{n})$ entries in expectation.
\begin{lemma}[Chart work]\label{lem:chart}
For each rule $A\to BC$, $M_B(n)M_C(n)\le 2n+1$. The average total
cost of constructing the designated charts is $O(n^{1.5})$.
\end{lemma}
\begin{proof}
By unambiguity,
concatenation is injective on $L_B\times L_C$, since a collision would
give two parses from $A$~\cite[Corollary 1]{KT69}.
Each pair has weight $p(u)p(v)=p(uv)$. When
$|u|,|v|\le n$ there are at most $2n+1$ output lengths, each of total
probability at most one. This proves the inequality and gives
$M_Y(n)\le\sqrt{2n+1}$ for the designated child $Y$.

For each descendant $X$ of $Y$, fix a derivation $Y\Rightarrow^*vXt$
with fixed terminal words $v,t$ and put $d=|v|+|t|$. Embedding
each $u\in L_X$ as $vut$ gives
\[
 s^{-d}M_X(n)\le M_Y(n+d)\le M_Y(n)+d.
\]
The last inequality uses the fact that each extra length adds at most one to
the mass. Hence $M_X(n)=O(\sqrt n)$. At any starting position the expected
number of matching $X$-spans is at most $M_X(n)$, by summing over lengths.
Summing over the $n$ positions and the finitely many roots and descendants,
then applying Lemma~\ref{lem:manacher}, proves the chart bound.
\end{proof}

\begin{lemma}[Search work]\label{lem:search}
The $\Search$ recursion takes $O(n)$ average time excluding the time
for each chart construction.
\end{lemma}
\begin{proof}
In the algorithm, a recursive call to $\Search(D,i,j)$ happens when
the left and the right of the range $[i,j)$ have been already parsed
by the charts. Let the left and right terminal words be $u$ and $v$ and
let the sum of their lengths $k = |u|+|v| = n-j+i$.

Among the computations for all possible length-$n$ input words, how many
times is each such $(u,D,v)$-call visited?
Let $z_D$ be one shortest nonempty word in $L_D$, not necessarily
a subword of the input. $z_D$ is determined solely by the grammar.
Due to the unambiguity, $u z_D v$ uniquely determines
one parse tree, and therefore, the word $u z_D v$ uniquely determines
the search history that reaches the call to $\Search(D)$ with the context
$(u,v)$. Here, a search history is the root-to-call path of ancestor
rules and completed sibling subtrees.
Starting at the root, the choice of whether to construct the chart
for the left or right child fixes the path that follows the search child.
Subtree lengths strictly decrease on this path, so it contains at most one
node of length $|z_D|$. This identifies the filled hole and recovers
the entire search history.

For fixed $D,k$, there are at most $s^{k+|z_D|}$ distinct $u z_D v$ words and
therefore at most that many histories. Visiting any one history requires
matching its prefix $u$ and suffix $v$ in $W_n$. For $k<n$ these positions
are disjoint, so the match has probability $s^{-k}$. The expected number of
visited histories for this $D,k$ is therefore at most $s^{|z_D|}$, a constant.
Summing over $k<n$ and $D$ gives $O(n)$ expected calls.
\end{proof}

\begin{lemma}\label{thm:mainlemma}
The algorithm takes $O(n^{1.5})$ expected time on a uniformly random word
of length $n$ and $O(n^2)$ time on every word of that length.
\end{lemma}
\begin{proof}
With the oracle, Lemmas~\ref{lem:chart} and~\ref{lem:search} give the
average time bound. On a fixed input, two calls with the same $(D,i,j)$
would have the same outside words and filler, hence the same history by
the injection in Lemma~\ref{lem:search}. Each history is visited at most
once, so there are $O(n^2)$ calls in the worst case. There are also
$O(n^2)$ chart entries. Hence the worst-case bound also follows.
\end{proof}

%%%%%%%%%%%%%%%%%%%%%%%%%%%%%%%%%%%%%%%%%%%%%%%%%%%%%%%%%%%%%%%%%%%%%%%%%%%

\section{Removing the Oracle}\label{sec:oracle}

We present two approaches to removing the mass oracle.
They differ in terms of implementation cost and/or practical
execution time, but in terms of asymptotic time complexity,
both implementations retain the expected $O(n^{1.5})$ bound.

\subsection{Interleaving Chart Computations}

The key observation of the first approach is that we do not need the exact
values of the masses. We only want to choose the child whose chart
can be constructed faster. To know which child runs faster, we simply run
the chart creation jobs for all right-hand side nonterminals in parallel,
and once one job of a rule finishes, abort the other one.

As a precomputation before calling $\Search(S,0,n)$,
start a resumable chart-and-index job for the whole input word for each
distinct binary-rule operand.
Run all jobs round-robin, including their calls to Manacher's algorithm.
Stop once every rule has at least one finished operand. Choose a finished
side for each rule, with a fixed tie rule, and discard unfinished jobs.
At length $n$, let $Y_r$ be an oracle-preferred operand for rule $r$.
Write $T_Y(w)$ for the cost of running the job for $Y$ to completion.
After $\max_r T_{Y_r}(w)$ rounds all rules are covered, and each round does
constant work. The expected maximum is at most the sum of the expectations
of the oracle-chosen jobs, which is $O(n^{1.5})$ by Lemma~\ref{lem:chart}.

A subtlety is that the use of completion times makes the orientation choice of
whether to chart the left or right child depend on the input word.
Pointwise, its search cost is bounded by the sum of the search costs for
\emph{all} fixed orientations on the same word. Lemma~\ref{lem:search}
bounds each expectation by $O(n)$, and the number of orientations is
at most $2^{|R|}$, a fixed constant determined only by the grammar, not
the input word.

\subsection{Computing the Mass Exactly}

The second approach is more direct.
The counting function of an unambiguous context-free language is well
studied and is efficiently computable.
Let $c_X(k)=|L_X\cap\Sigma^k|$ and define the formal power series
 $C_X(z)=\sum_{k\ge0}c_X(k)z^k$.
If $t_A$ counts the rules $A\to a$, unambiguity gives
\[
 C_A(z)=t_Az+\sum_{A\to BC}C_B(z)C_C(z).
\]
This system defines algebraic power series~\cite{CS63}. Compare the integer
scores $b_X(n)=s^n M_X(n)$. Since $b_X(0)=0$ and
$b_X(k)=s\,b_X(k-1)+c_X(k)$, their generating function is $C_X(z)/(1-sz)$.
It too is algebraic, so its coefficients satisfy an effectively computable
fixed-order recurrence with polynomial coefficients
\cite{Stanley80}. Precompute it and the initial values past
any integer zeros of its leading coefficient.

To obtain $b_X(n)$ alone, multiply the fixed-size recurrence matrices by
balanced binary splitting, using a common integer denominator per matrix
\cite{Mezzarobba12}. Each factor has $O(\log n)$-bit entries, and each
product-tree level has total operand length $O(n\log n)$ bits. Fast integer
multiplication gives $\widetilde O(n)$ bit operations overall, where the
tilde hides logarithmic factors. Comparing the resulting scores realizes the oracle.

\subsection{Main Theorem}

\begin{theorem}\label{thm:main}
For every fixed unambiguous context-free grammar there is a deterministic
parser that returns the unique parse tree or rejects. It takes $O(n^{1.5})$
expected time on a uniformly random word of length $n$ and $O(n^2)$ time
on every word of that length.
\end{theorem}
\begin{proof}
By Lemma~\ref{thm:mainlemma} and the oracle removal explained above.
\end{proof}

%%%%%%%%%%%%%%%%%%%%%%%%%%%%%%%%%%%%%%%%%%%%%%%%%%%%%%%%%%%%%%%%%%%%%%%%%%%

\section{Concluding Remarks}\label{sec:conc}

The argument used in Lemma~\ref{lem:chart} for bounding the mass for
each nonterminal has further implications beyond the lemma.
In fact, the proof establishes a stronger
inequality $M_B(n)M_C(n)\le M_A(2n)$. This immediately gives us a
finer-grained average-case time bound: $O(n \sqrt{M_S(2n)})$.
For instance, for a language with $O(\sqrt{n})$ mass, 
the algorithm presented in this paper runs in expected $O(n^{1.25})$ time.
Note that the exponent of the asymptotic order of the mass function is
always a dyadic number~\cite{BD15}.

Another class of languages that admits lower average-time complexity
consists of the languages generated by linear unambiguous grammars
where the right-hand side of each rule (before normalizing to Chomsky
normal form) contains at most one nonterminal. This is because all but
one child of each rule has constant mass, which implies $O(n)$ bound
for the expected total parsing time.

An interesting observation can be made for challenging cases as well:
the hardest rules splitting the mass evenly to $\Theta(\sqrt{n})$
times $\Theta(\sqrt{n})$ appear at most once in each parse tree. Otherwise
the total mass $M_S(n)$ would exceed $n$.
If we can handle this single difficult rule in sub-$O(n^{1.5})$ time, this would
yield an improved upper bound for general unambiguous context-free grammars.
We leave further applications of this technique to future work.

%%%%%%%%%%%%%%%%%%%%%%%%%%%%%%%%%%%%%%%%%%%%%%%%%%%%%%%%%%%%%%%%%%%%%%%%%%%

\paragraph{AI Assistance Disclosure:}
This work was developed with substantial assistance from OpenAI's
ChatGPT (GPT-6, accessed November 2026). In particular, it generated
the proof of Lemma~\ref{lem:search}, the odd-cycle example in
Section~\ref{sec:intro}, and the observation of the single
hardest split in Section~\ref{sec:conc}. The author verified and revised
the proof and takes full responsibility for the paper.

\bibliographystyle{alpha}
\bibliography{mybib}

\end{document}